\documentclass[pra,twocolumn]{revtex4}
\usepackage{mathrsfs,latexsym,amsmath,psfrag,exscale,epsfig}
\newcommand{\eq}[1]{Eq.~(\ref{#1})}
\newcommand{\fig}[1]{Fig.~\ref{#1}}
\begin{document}
\title{\large\bf Enhanced ionization in small rare gas clusters}
\author{Christian Siedschlag}
\author{Jan M. Rost}
\affiliation{
 Max-Planck-Institute for the Physics of Complex Systems,
N\"othnitzer Str. 38, D-01187 Dresden, Germany}
\date{\today}

\begin{abstract}
A detailed theoretical investigation of rare gas atom clusters under
intense short laser pulses reveals that the mechanism of energy
absorption is akin to {\it enhanced ionization} first discovered for
diatomic molecules.  The phenomenon is robust under changes of the
atomic element (neon, argon, krypton, xenon), the number of atoms in
the cluster (16 to 30 atoms have been studied) and the fluency of the
laser pulse.  In contrast to molecules it does not dissappear for
circular polarization.  We develop an analytical model relating the
pulse length for maximum ionization to characteristic parameters of 
the cluster.
\end{abstract}
\maketitle
\section{Introduction}
Building the bridge between atomic and solid state physics, cluster
physics has become a vivid research field of its own.  While the
static properties of clusters are by now well understood, there remain
many open problems concerning the dynamics of clusters under external
perturbations.  In the case of weak perturbations, linear response
theory has proven to be a valid tool for the investigation of
dynamical properties \cite{cal3}; but with increasing strength of the
perturbation the description of the cluster evolution becomes more and
more involved \cite{cal2}.

On the other hand, experimental studies of (mostly) rare gas cluster
interaction with highly charged projectiles as well as with short,
intense laser pulses have produced a number of interesting results
calling for an explanation.  
Experiments exploring the interaction of rare gas clusters with intense laser light 
have shown a big increase of energy
absorption compared to the single atom case
\cite{koe1,dit1,zwei1,lez1,pur1,spri1}.  When irradiated with a
$10^{15} \rm{W/cm}^2$ femtosecond laser whose wavelength is in the
optical regime, one observes, depending on the cluster size and the
atomic element, ionic charge states of up to 40.  These high charge
states let the fragmenting ions gain an enormous amount of kinetic
energy.  The most spectacular example for this highly energetic
process has certainly been the recent experimental observation of
nuclear fusion in a cluster \cite{dit4}.

Here, we focus on clusters of some 10 atoms.  We have developed a
model containing the essential features of the interaction between the
cluster and the laser field.  It will be shown that energy absorption
from the laser pulse proceeds through a mechanism originally
discovered for diatomic molecules ({\em enhanced ionization} (ENIO)
\cite{zuo3,sei1}), whose generalization for the case of small clusters
will be presented.

The paper is organized as follows: after introducing the numerical
model and comparing it to other types of simulations already existing
in the literature in section II, we investigate the dependence of
energy absorption and ionization yield on the pulse length in a series
of clusters in section III. From the results of these calculations
 a generic behavior emerges which can be explained by invoking the
above-mentioned enhanced ionization mechanism as explained in section
IV. We give strong evidence that this mechanism should play an
important role in the laser-cluster interaction over a large range of
parameters detailed in section V. Finally we condense our picture of
the ionization process into a simple analytical expression which
quantifies the role of the experimentally accessible variables like
cluster size or atomic element on the process of energy absorption in
section VI. The last section VII summarizes our work.  Atomic units
are used if not stated otherwise.

\section{The cluster model}
\subsection{Theoretical formulation and numerical implementation}
  Since the dimension of the problem is far too high to allow for an
 exact quantum mechanical treatment we have formulated a model to
 describe the dynamics of rare gas clusters in strong laser fields. 
 We resort to a classical treatment, with a few but essential quantum
 mechanical elements.

Initially, before the onset of the laser pulse, we find the
equilibrium ground state configuration of the cluster with
Lennard-Jones interaction between the neutral cluster atoms.  The
global minima for this potential are readily available.  The electrons
are assumed to be localized at the nuclei.

After fixing the initial shape of the cluster, we start  the
time evolution switching on the laser pulse \cite{footnote1}. 
For the electrons, this evolution consists of two parts:
firstly, the modeling of the bound state and the process of ionization
from this state; secondly, the propagation after being ionized from an
atom.  We will refer to the first process as {\em inner ionization},
in contrast to the {\em outer ionization}, which has the effect that
an electron leaves the cluster.  Inner ionization contains processes
beyond classical mechanics, while the subsequent propagation and
(eventually) outer ionization is described classically via integration
of Newton's equations.

When irradiating a cluster with intense laser light, two processes can
lead, at least in principle, to inner ionization: {\em field
ionization} and {\em electron impact ionization}.  In the case of
field ionization, the electric field inside the cluster (initially
only the laser field, later the combined field of laser, ions and
electrons) leads to a lowering of the potential barriers, so that an
electron can leave its mother atom via tunneling \cite{amm1} or even
{\em over-the-barrier} \cite{bau2}.  Electrons which are already inner
ionized, but not yet outer ionized, can further lead to electron
impact ionization.  This mechanism was shown to play almost no role in
small clusters \cite{ishi1}, as the average free path length with
respect to electron impact ionization is much larger than the cluster
radius.  For this reason we only consider field ionization in our
model.

The model is implemented as follows.  Before the onset of the pulse,
the electrons of the cluster are assumed to be localized at the atomic
positions; the bare nucleus and all the electrons of an atom are
treated as one neutral classical particle.  It is only later that the
electrons (through inner ionization) are born as separate classical
particles.  Hence, the number of particles in our simulation changes
with time.

\begin{figure}
\centering
\psfrag{xtitle}[][][1]{r}
\psfrag{ytitle}[][][1]{energy}
\psfrag{kurve1}[l][][0.9]{V(r)}
\psfrag{kurve2}[l][][0.9]{$E_n$}
\psfrag{r1}[][b][1]{$r_1$}
\psfrag{r2}[][b][1]{$r_2$}
\epsfig{file=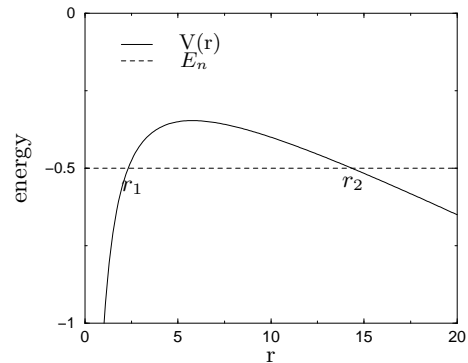,width=0.7\columnwidth}
\caption{Tunneling from a bound state with energy $E_n$ in the potential $V(r)$}
\label{tunnelbild}
\end{figure}

Two classical charged particles at positions $\vec{r}_1$ and $\vec{r}_2$
interact via the potential
\begin{align}
\label{softcore}
V(\vec{r}_1,\vec{r}_2)=Z_1Z_2\left(\left|\vec{r}_1-\vec{r}_2\right|^2
+a_{Z_1}+a_{Z_2}\right)^{-1/2},
\end{align}
where $Z_1$ and $Z_2$ are the charges of the two particles. The 
$a_{Z_i}$ are
softcore parameters, which help to regularize the Coulomb singularity. We are
using a $Z$-dependent $a$, so that the depth of the atomic potential can be
adjusted to the current binding energy. To determine $a_{Z}$, we 
proceed as follows:
\begin{itemize}
\item[--]{For electrons, $a$ has been chosen to be 0.1, i.e.\  $a_{-1}=0.1$}
\item[--]{For an electron which is localized in the minimum of the potential
of an atom with charge $Z$, we demand
\begin{align}
\frac{-Z}{\sqrt{a_{Z}+a_{-1}}}=E_{\rm bind}(Z)+\epsilon 
\end{align}
with  a small positive parameter $\epsilon=0.01$. 
Then we can solve for $a_{Z}$
\begin{align}
\label{avonz}
a_{Z}=\frac{Z^2}{E_{\rm bind}^2+2\epsilon E_{\rm 
bind}+\epsilon^2}-a_{-1}\,.
\end{align}
}
\end{itemize} 
At every time step $dt$, we calculate for each atom the ionization probability
of the outermost electron in the following fashion: if $\vec {B}_j$ is the
total electric field at the position $\vec{R}_j$ of atom $j$,
\begin{align}
\vec{B}_j=\vec{\nabla}_{R_j}\left[\sum_{i\neq j}
\frac{Z_i}{\sqrt{(\vec{R}_j-\vec{R}_i)^2+a_{Z_i}+0.1}}\right]+\vec{\epsilon}_p f(t),
\end{align} 
with the polarization vector $\vec{\epsilon}_p$ of the laser field, we calculate
the  tunneling integral in the direction of $\vec{B}$:
\begin{align}
I(t)=\exp\left[-2\int_{r_1}^{r_2}\sqrt{2(V(r)-E_n)} \ dr \right].
\end{align}
The role of $r_1$ and $r_2$ is shown in \fig{tunnelbild} where $V(r)$
contains the Coulombic potential terms and those coming from the laser field. The energy level $E_n$
is defined as
\begin{align}
E_n:=E_n^{\rm Atom}+V(0)-\frac{Z_{\rm Atom}+1}{\sqrt{a_{Z_{\rm Atom}}+0.1}}\,.
\end{align}
Hence, the atomic energy level is shifted by the surrounding charges and the
laser potential; the potential of the atom from which the electron will be
ionized has to be subtracted because its influence is already contained in
$E_n^{\rm Atom}$. The positions $r_1$ and $r_2$, where $E_n$ crosses the potential
curve $V(r)$, are determined numerically; the search for $r_2$ is continued
until $I(t)<10^{-10}$. If we find a position $r'$ along the direction of
$\vec{B}$ with $dV/dr|_{r=r'}=0$ and $E_n>V(r)$ for $r\leq r'$,
over-the-barrier ionization is possible. In this case we simply put 
$I(t)=1$.
The tunneling rate $w(t)$ is then the tunneling probability $I(t)$
multiplied with the frequency of the electron hitting the potential
barrier.  In a semiclassical picture, this frequency is just the
inverse of the Kepler period $T_n$ belonging to an orbit with binding
energy $E_n$:
\begin{align}
T_n=\pi (Z_{\rm Atom}+1)/(2E_n^3)^{-1/2}. 
\end{align}
Hence, the tunneling rate is 
\begin{align}
w(t)=I(t)/T_{n}.
\end{align}

The tunneling probability over a unit of time $dt$ is $P(t)=w(t) \, dt$.  
By comparison with a random number $z$ (is $P(t)>z$?)  we decide
if the electron in question tunnels.  If so, we place the electron,
which now becomes a classical particle, outside the potential barrier
as close to $r_2$ as possible, with the exact position and momentum of
the electron determined by conservation of the total energy
\cite{footnote2}.  If the ionization happens to be
over-the-barrier, we put the electron on top of the barrier,
where $dV/dr=0$.  The atomic charge is raised by 1 and the next
virtual electron is allowed to tunnel.

The particles are classically propagated by integrating Newton's
equations of motion.  We have used a symplectic integrator \cite{cha3}
with a time step of $dt=0.1$.

\subsection{A typical run}
Although later we will use a Monte Carlo ensemble to calculate
experimentally accessible observables, for a qualitative understanding
of the phenomena it is sufficient to have a closer look on a single
event, since the overall behavior of the ensemble members is quite
similar.  As an example we consider a Ne$_{16}$-cluster.  The applied
pulse has a peak intensity of $I=10^{15} \ {\rm W/cm^2}$, a frequency
of $\omega =0.055$ a.u. (780 nm) and it extends over 20 cycles, so
that the pulse length is approx.  $T\approx 55$ fs.  We chose a
$\sin^2$-function for the pulse envelope, i.e. the pulse is of the
form
\begin{align}
\label{Puls}
f(t)=F\sin^2\left({\pi t}/{T}\right)\sin(\omega t) \hspace{2 cm} (0\leq t\leq T).
\end{align}

\begin{figure}
\begin{center}
\psfrag{xtitle}[][][1]{t [a.u.]}
\psfrag{ytitle}[][][1]{Number of classical particles}
\psfrag{index}[][][1]{a)}
\epsfig{file=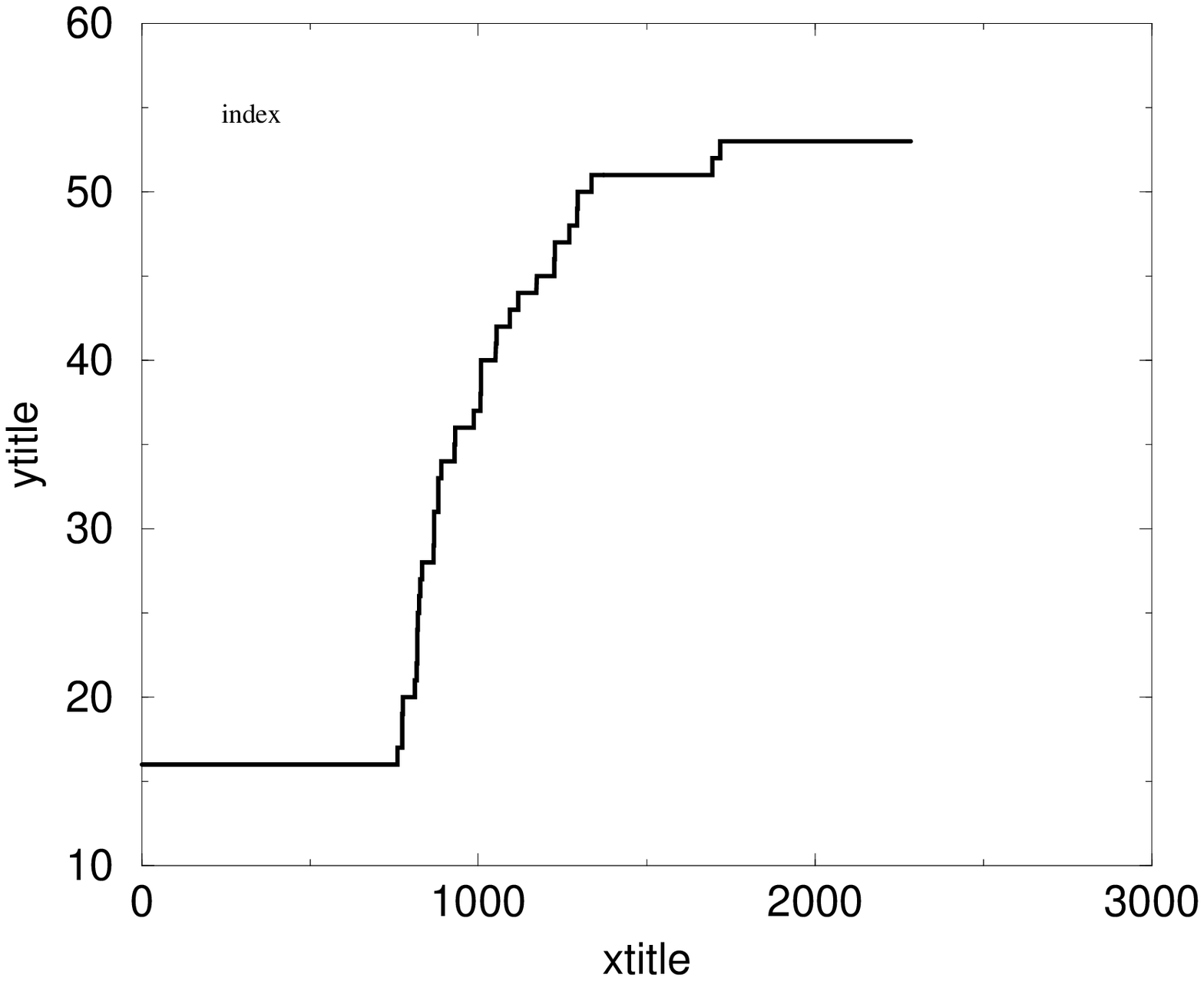,width=0.7\columnwidth}
\psfrag{ytitle}[][][1]{Cluster charge}
\psfrag{index}[][][1]{b)}
\epsfig{file=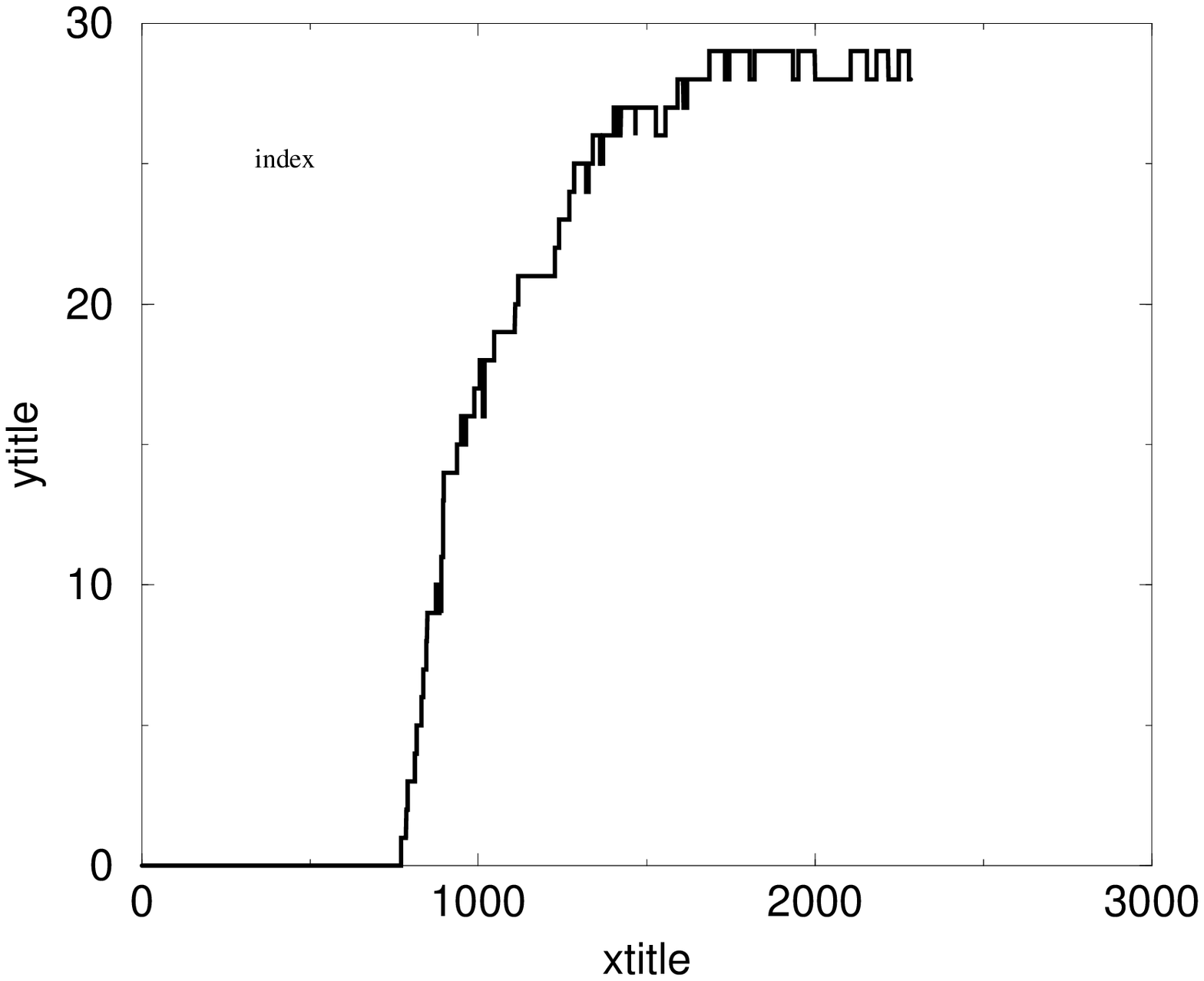,width=0.7\columnwidth}
\caption{Number of classically treated particles (a) and   total charge 
of the cluster (b).}
\label{einzelrun1}
\end{center}
\end{figure}
\fig{einzelrun1} shows the total number of classically treated particles
and the cluster charge, i.e. the number of electrons which have left the
cluster, as a function of time. After approximately $750$ fs the intensity
of the laser is sufficiently high for the first inner ionization event,
followed by a rapid increase of the number of classical particles as well as
the cluster charge. Obviously, the ionization of the first few electrons leads
to an  `avalanche effect': the inner ionized electrons create a strong electric
field inside the cluster, which helps to inner ionize further electrons (this
is reminiscent of the {\em ionization ignition} mechanism \cite{ros1}). 

\begin{figure}
\begin{center}
\psfrag{xtitle}[][][1]{t [a.u.]}
\psfrag{ytitle}[][][1]{absorbed energy [eV]}
\psfrag{index}[][][1]{a)}
\epsfig{file=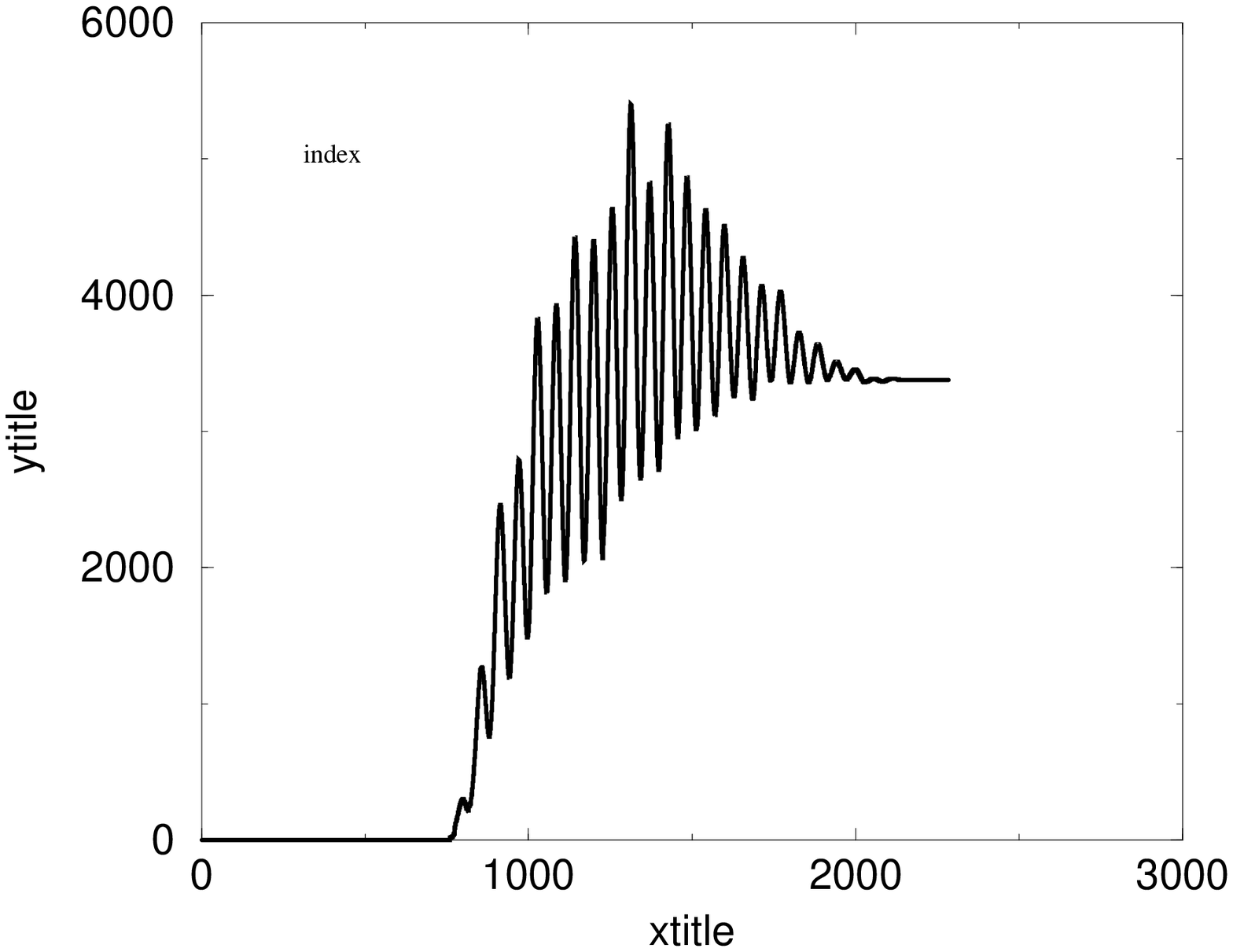,width=0.7\columnwidth}
\vspace{1 cm}
\psfrag{ytitle}[][][1]{mean interionic distance [a.u].}
\psfrag{index}[][][1]{b)}
\epsfig{file=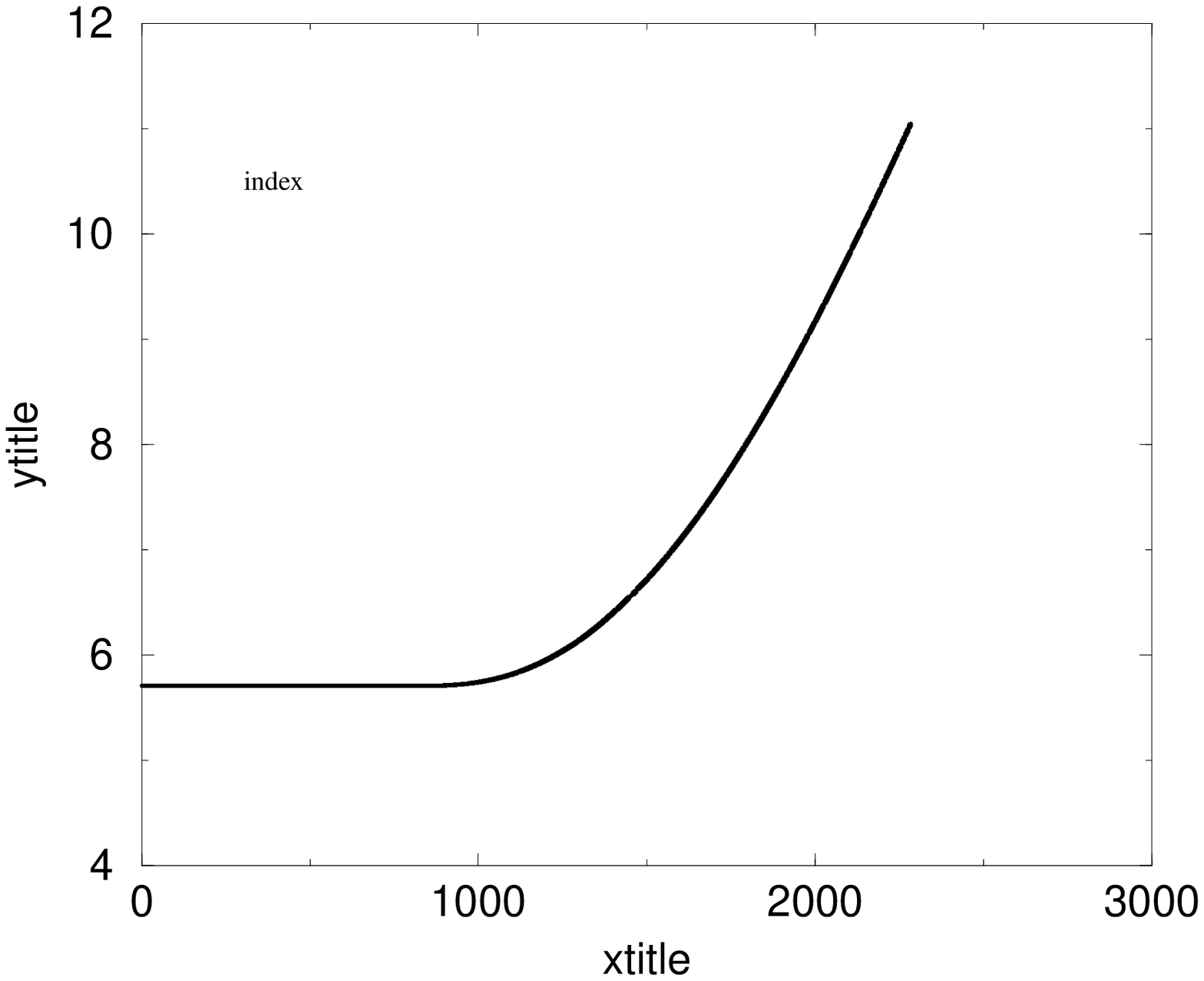,width=0.7\columnwidth}
\caption{Absorbed energy (a) and mean interionic distance (b) (see \eq{R})
as a function of time.}
\label{einzelrun2}
\end{center}
\end{figure}

Fig.  \ref{einzelrun2} shows the energy absorbed by the cluster and
the mean interionic distance, again as a function of time.  If
$\mathscr{K}$ is the set of nuclei with mass $M$, $\mathscr{E}$ the
set of inner ionized electrons (mass $m$) already treated classically
and $\mathscr{G}$ the set of electrons which are still bound, the
cluster energy is defined as
\begin{align}
\label{Energie}
\sum_{i\in\mathscr{K}}&\frac{P_i^2}{2 M}+\sum_{i\in\mathscr{E}}\frac{p_i^2}{2
m}+\sum_{i,j \in\mathscr{K}\cup\mathscr{E}} V_{ij}+\sum_{i\in\mathscr{G}}
E^{\rm bind}_i \\
&+\sum_{i\in\mathscr{K}}Z_i\left(\vec{\epsilon}_p\cdot\vec{R}_i\right)f(t)
- \sum_{i\in\mathscr{E}} \left(\vec{\epsilon}_p\cdot\vec{r}_i\right) \ 
f(t)\,.
\end{align}
 The absorbed energy is the difference of the total energies before and
 after the laser pulse.  As we can see, this rather small cluster can
 already absorb  a considerable amount of energy.  The oscillations are due to
 the ponderomotive potential and have no direct influence on the net
 energy absorption.  As the cluster gets charged, it begins to expand,
 i.e.\ the mean interionic distance will increase.  For a cluster
 consisting of $N$ atoms, it is defined as
\begin{align}
\label{R}
R(t) = \left(\frac{1}{N}\sum_{i=1}^N \min_{i \neq j} \{|\vec R_{i}-\vec
R_{j}|^2\}\right)^{1/2}.
\end{align}

At the intensity used here, the cluster disintegrates completely, i.e.
we observe only atomic fragments after the pulse.  Note that the
expansion of the cluster takes place adiabatically compared to the
time scales of the laser frequency and the electronic motion, but on
the same time scale as the pulse length.  Hence, it is possible to
explore radius-dependent properties of the cluster by varying the
pulse length.

\subsection{Comparison to other models}
Similar models for the theoretical description of small rare gas
clusters in strong laser fields have been proposed by other authors. 
The common feature is the use of classical mechanics as the basic
ingredient of the description; the main differences lie in the
treatment of inner ionization.

\subsubsection{The onion model}
Following its name this model \cite{ros1}, as ours, treats the weakest
bound electron per atom or ion as the active electron which can be
inner ionized.  However, tunneling is not included.  Instead, the
active electron circles its mother ion on a Kepler orbit with an
energy corresponding to the quantum mechanical binding energy.  This
orbit is then deformed in time by the laser field and the surrounding
charges, and inner ionization can take place, which is defined by
reaching a certain distance $r_c$ from the mother ion.  If the
electron exceeds this distance, it is assumed to be inner ionized, and
the next electron is put onto a (now deeper bound) Kepler orbit.

In common with our model, inner ionization takes place sequentially,
and the already inner ionized electrons are propagated classically. 
By neglecting any tunneling contributions, the first inner ionization
will take place a certain time $\delta t$ later than in our case.  As
we will see, this delay can have a significant influence on the
subsequent dynamics of the cluster rendering a fully classical
treatment as in the onion model problematic.

\subsubsection{Tunneling via the Landau rate}
In this ansatz \cite{dit6} the tunneling mechanism is taken into
account, but the tunneling integral is not calculated explicitely. 
Instead, the time dependent tunneling probability is estimated using
the Landau tunneling rate \cite{lan2}, which has been established for
the case of an atom in a quasistatic laser field.  For this purpose,
the total electric field (laser + electrons + ions) at the position of
an atom is used.  For inner ionization to happen in this model the
field strength has to be strong enough at just one point, namely, the
position of the atom.  Hence, an electron which comes by chance close
to an ion will create {\em locally} such a strong field that
ionization can hardly be avoided.  In our model, in contrast, the
entire environment of an atom must be suitable for ionization. 
Consequently, we get lower ionization rates than in \cite{dit6} but in
better agreement with \cite{ishi1}.  \subsubsection{Using the full
Coulomb potential} In a third model \cite{ishi1}, the full tunneling
integral is calculated at each time step as in our case.  Moreover,
the authors do not use a softcore potential but employ the full
Coulomb potential and regularize the singular equations of motion. 
The use of the full Coulomb potential may, at first glance, seem to be
an advantage.  However, it leads to the problem of classical
autoionization: a classical electron in a Coulomb potential can be
arbitrarily deeply bound, since there is no uncertainty relation.  The
corresponding energy gained is available as kinetic energy for the
ionization of other electrons.  In our model, this unphysical behavior
is avoided by requiring the minimum of the softcore potential to
coincide with the quantum mechanical binding energy.

In \cite{ishi1} a recapture mechanism was built in to account for this
problem, which can transform already classically treated electrons
back into a virtual bound state existence.  Not to mention the
peculiarities which arise when defining the exact conditions for
recapture in a many-particle environment, the numerical expenditure is
higher than in our case.  Nevertheless, the ionization yields calculated
with this model agree quite well with our results.

\section{Absorption properties for different pulse lengths}
\label{meiwes}
To investigate how the expansion of a cluster during the interaction with a
strong laser pulse influences its absorption behavior, we have calculated the
absorbed energy and the average ionic charges after the interaction for various pulse
lengths. To keep the amount of energy delivered by these pulses fixed, we
demand the fluency to be constant, i.e.
\begin{align}
\label{normierung}
E(T):=\int_0^T f^2(t) \ dt = \rm{const.}
\end{align}
For a pulse of the shape of \eq{Puls}, we obtain
\begin{align}
E(t)={3F^2T}/{16}.
\end{align}
For the reference pulse, we chose the parameters already used in the single
run from the previous section: $F=0.16$ a.u., $\omega=0.055$ a.u. and a pulse
length of 20 cycles. Shorter pulses have a higher maximum field strength
and longer pulses a lower maximum field strength, respectively. The results
were obtained by averaging over a Monte Carlo ensemble consisting of 20
clusters.
\subsection{The light Ne$_{16}$ cluster: almost atomic behavior}
\label{versch_elem}
\begin{figure}
\begin{center}
\psfrag{xtitle}[][][0.9]{T [a.u.]}
\psfrag{ytitle}[][][0.9]{absorbed energy [eV]}
\epsfig{file=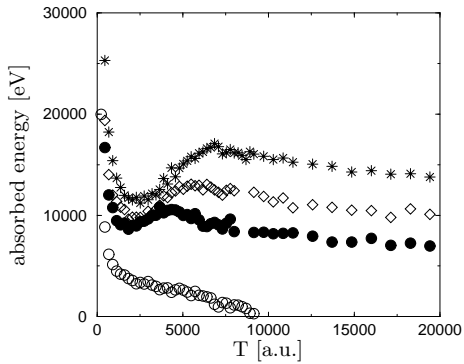,width=0.7\columnwidth}
\caption{Energy absorption of Ne$_{16}$ ($\circ$), 
 Ar$_{16}$ ($\bullet$),  Kr$_{16}$ ($\diamond$), and
 Xe$_{16}$ ($\star$) for different pulse
lengths (see text).}
\label{meiwes.en}
\end{center}
\end{figure}

\begin{figure}
\begin{center}
\psfrag{xtitle}[][][0.9]{T [a.u.]}
\psfrag{ytitle}[][][0.9]{average atom charge}
\epsfig{file=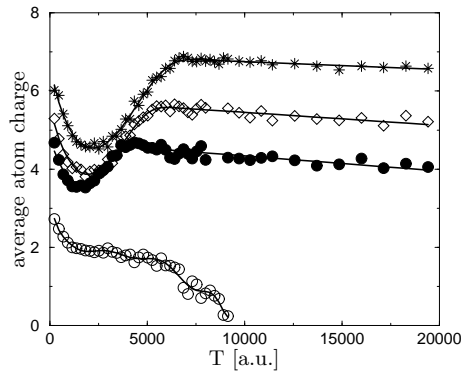,width=0.7\columnwidth}
\caption{Average atom charge as a function of pulse lengths for 
different clusters, see \fig{meiwes.en}.  The solid lines are to guide the eye.}
\label{meiwes.charge}
\end{center}
\end{figure}

The first cluster we consider is Ne$_{16}$. The absorbed energy and the
average atomic charges as a function of pulse length under the constraint of
\eq{normierung} are 
shown in Figs.\ \ref{meiwes.en} and \ref{meiwes.charge}, respectively. 

Both curves decrease monotonically with increasing pulse length.  The
average atomic charge exhibits a plateau for pulse lengths around
$T=3000$ a.u..  At this point, we can already suspect that the plateau
might have its origin in the expansion of the cluster during the
pulse.  However, in the case of Ne$_{16}$ its effect seems to be
rather small.  In fact, Neon clusters of this size behave more like
atoms.  This changes for heavier rare gases.

\subsection{Generic cluster behavior in Ar$_{16}$}
The absorbed energy (\fig{meiwes.en}) and average atomic charge
(\fig{meiwes.charge}) for an Ar$_{16}$ cluster decrease
monotonically for very short pulse lengths only.  Both quantities start to rise
again for $T\approx 2000$ a.u., reaching a maximum at an optimal pulse
length of $T\approx 4500$ a.u. (we will refer to this optimal pulse
length as $T^*$ subsequently) before they ultimately decrease.  The
average charge state reached at $T=T^*$ is around 4.5, which is
considerably higher than what one would expect from a single Argon
atom.

The maximum is much more pronounced for the ionization yield than for the
absorbed energy for two reasons. Firstly, for shorter pulses
the mean internuclear distance just after the pulse will in general be smaller
than for longer pulses, so that the Coulomb explosion energy will increase
with decreasing pulse length. Secondly, the ionized electrons, which can be
considered to be quasi-free, acquire a higher kinetic energy for shorter
pulses, since the intensity is higher than for longer pulses. These two
effects wash out  the minimum in the curve for the absorbed
energy and  decrease the contrast between minimum and maximum.

We will postpone the further discussion and explanation of this structure 
after we have taken a look at the corresponding results for
Kr$_{16}$ and Xe$_{16}$.
 

\subsection{Same qualitative behavior as for Argon: Kr$_{16}$ and Xe$_{16}$}
Qualitatively, the behavior for Kr$_{16}$ as well as for Xe$_{16}$ is
the same as for Ar$_{16}$: the energy absorption and the average
atomic charge peak for a certain pulse length $T^*$, which depends on
the type of the cluster.  The average atomic charge increases with the
atomic mass, reaching an average of almost $7^+$ for the Xe$_{16}$
cluster; this is of course due to the decreasing binding energies of
the electrons in the heavier elements.  The value of $T^*$ is changing
for different clusters; we will come back to this point later, after
we have understood the reason for the existence of $T^*$.

\section{Calculations with fixed nuclei}
\label{static}
The existence of an optimal pulse
length $T^*$ can be related to an optimal cluster geometry with 
critical  radius $R^*$, which maximizes
the energy absorption (and also the ionization) of a small rare gas
cluster in a strong field. Given the existence of such a critical radius,
which should be larger than the equilibrium radius $R_0$, the occurrence of
$T^*$ can be readily explained.

For very short pulse lengths, the cluster has almost no time to expand during
the pulse, so that the critical radius will be reached only well after the
pulse is already switched off. For longer and longer
pulses, the cluster will reach $R^*$ at earlier and earlier times. For a
certain pulse length $T^*$, the time of reaching $R^*$ will roughly
coincide with the maximum of the pulse, which leads to optimal absorption. If
the pulses are becoming even longer, $R^*$ will be reached already before
the maximum of the pulse {\em and} the maximum intensity is decreasing due to
the energy normalization (\eq{normierung}). Both effects lead to a decrease
in energy absorption as well as in the average atomic ionization for
$T>T^*$.

\begin{figure}
\begin{center}
\psfrag{xtitle}[][][0.9]{$R/R_0$}
\psfrag{ytitle}[][][0.9]{cluster charge}
\epsfig{file=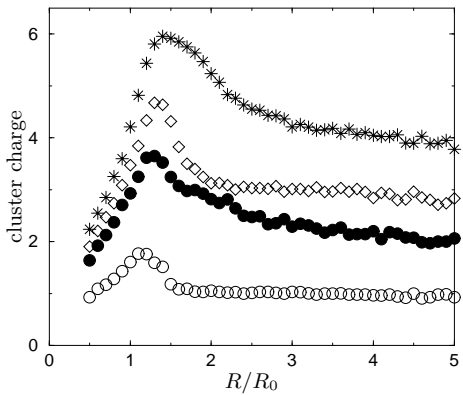,width=0.7\columnwidth}
\caption{Cluster charge, calculated with fixed nuclei, as a function of the
mean interionic distance (see \eq{R}, in units of the equilibrium mean
interionic distance $R_0$) for Ne$_{16}$ ($\circ$), Ar$_{16}$ ($\bullet$),
Kr$_{16}$ ($\diamond$) and Xe$_{16}$ ($\star$).}
\label{static.charge}
\end{center}
\end{figure}

What remains to be shown is that the critical radius $R^*$ really
exists and to explain its origin.  To this end, we have calculated the
cluster ionization yield for {\em different}, but {\em fixed} cluster
radii.  We have accomplished this by applying a scaling transformation
\begin{align}
\vec{R}_i^0 \Rightarrow \lambda \vec{R}_i^0
\end{align}
to the atomic positions, with $\lambda=1$ corresponding to the ground
state configuration.  The pulse we use is the reference pulse for
the calculations of the last section, i.e. of the form \eq{Puls} with
$F=0.16$ a.u., $\omega=0.055$ a.u. and 20 cycles  length.  The
results of these calculations are shown in \fig{static.charge}.
For all four clusters under consideration, we observe the existence of a
critical radius $R^*$ \cite{footnote3}, which is larger than the
equilibrium one. This means that it will be reached during the expansion the
clusters undergo when irradiated with intense laser light. The results of the
previous section thus can indeed be explained by the existence of
$R^*$.
The question remaining to be settled is: which mechanism is responsible for
$R^*$? In principle, two possibilities are available: first, it could be a
resonance effect where the electrons inside the cluster oscillate at a certain
characteristic frequency which would coincide with the laser frequency at a
certain cluster size. This kind of mechanism is well known from the plasmon
resonance in metal clusters \cite{eka1}; being originally a weak-field-concept,
the plasmon has been claimed to play an important role also in the strong
field regime \cite{koe1}. However, one needs a delocalized electron cloud to
create a plasmon resonance; in rare gas clusters, this condition is not
fulfilled.  

\begin{figure}
\psfrag{xtitle}[][][1]{electronic coordinate}
\psfrag{kurve1}[b][][0.9]{$R$=12 a.u.}
\psfrag{kurve2}[b][][0.9]{$R$=8 a.u.}
\psfrag{kurve3}[b][][0.9]{$R$=4 a.u.}
\begin{center}
\epsfig{file=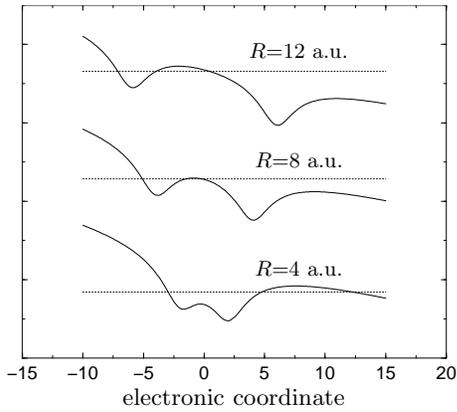,width=0.7\columnwidth}
\caption{Schematic potential curves and the upper
energy level 1$\sigma_+$ of a diatomic molecule
for different internuclear distances.}
\label{tunnelh2sketch}
\end{center}
\end{figure}

The second mechanism would be a generalization of a concept first discovered
for linear, diatomic molecules \cite{zuo3,post2,sei1}, called {\em enhanced
ionization } (ENIO). It can be qualitatively explained by looking at the potential
curve of such a molecule exposed to a quasistatic electric field (see
\fig{tunnelh2sketch}): the upper energy level of the two levels
$1\sigma_+$ and $1\sigma_-$, which emanate from the bonding and the antibonding
molecular orbital when an electric field is switched on, will lie above the
inner potential barrier but below the outer potential barrier when the
internuclear distance $R$ is rather small. On the other hand, when $R$ is
rather large, the level will lie below the inner barrier but above the outer
barrier (using the terms introduced in the previous section, we can say that
{\em inner ionization} is easier than {\em outer ionization} for small $R$ and
vice versa for large $R$). For an intermediate value of $R$, typically around
6-8 a.u., the interplay between inner and outer ionization will lead to a
maximum in the ionization rate.

This mechanism has been shown to be operative not only in linear molecules,
but also in triatomic molecules of triangular shape \cite{ban2,kaw1}. In this
case, the simple picture of \fig{tunnelh2sketch} is already slightly
distorted, and it is more appropriate to think of enhanced ionization in terms
of an optimal balance between inner and outer ionization, which makes the
generalization of the mechanism to a true many-body system like a cluster 
much easier.

\begin{figure}
\psfrag{xtitle}[][][1]{$R/R_0$}
\psfrag{ytitle}[][][1]{atomic charge}
\begin{center}
\psfrag{index}[][][1]{a)}
\epsfig{file=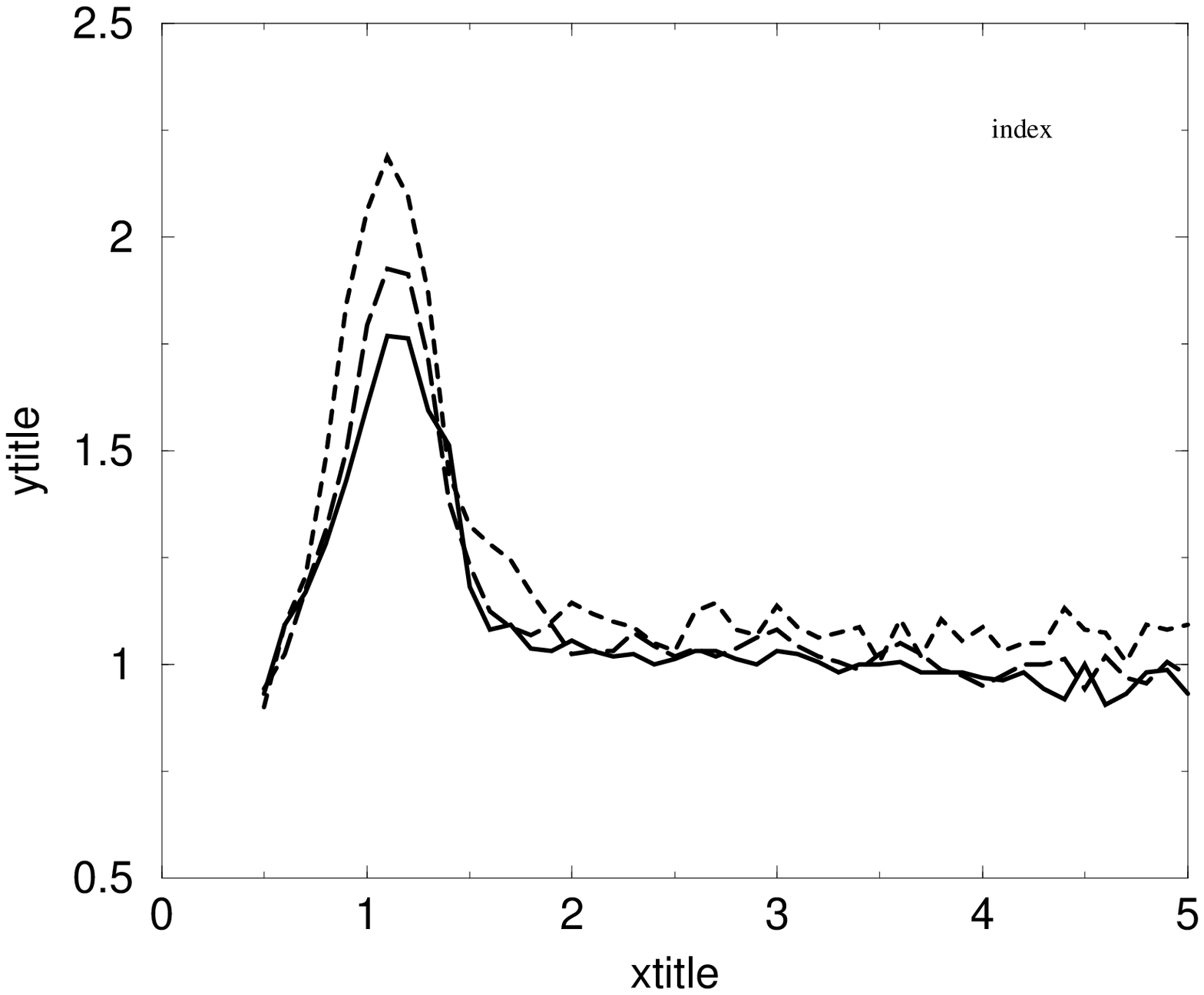,width=0.7\columnwidth}
\psfrag{index}[][][1]{b)}
\epsfig{file=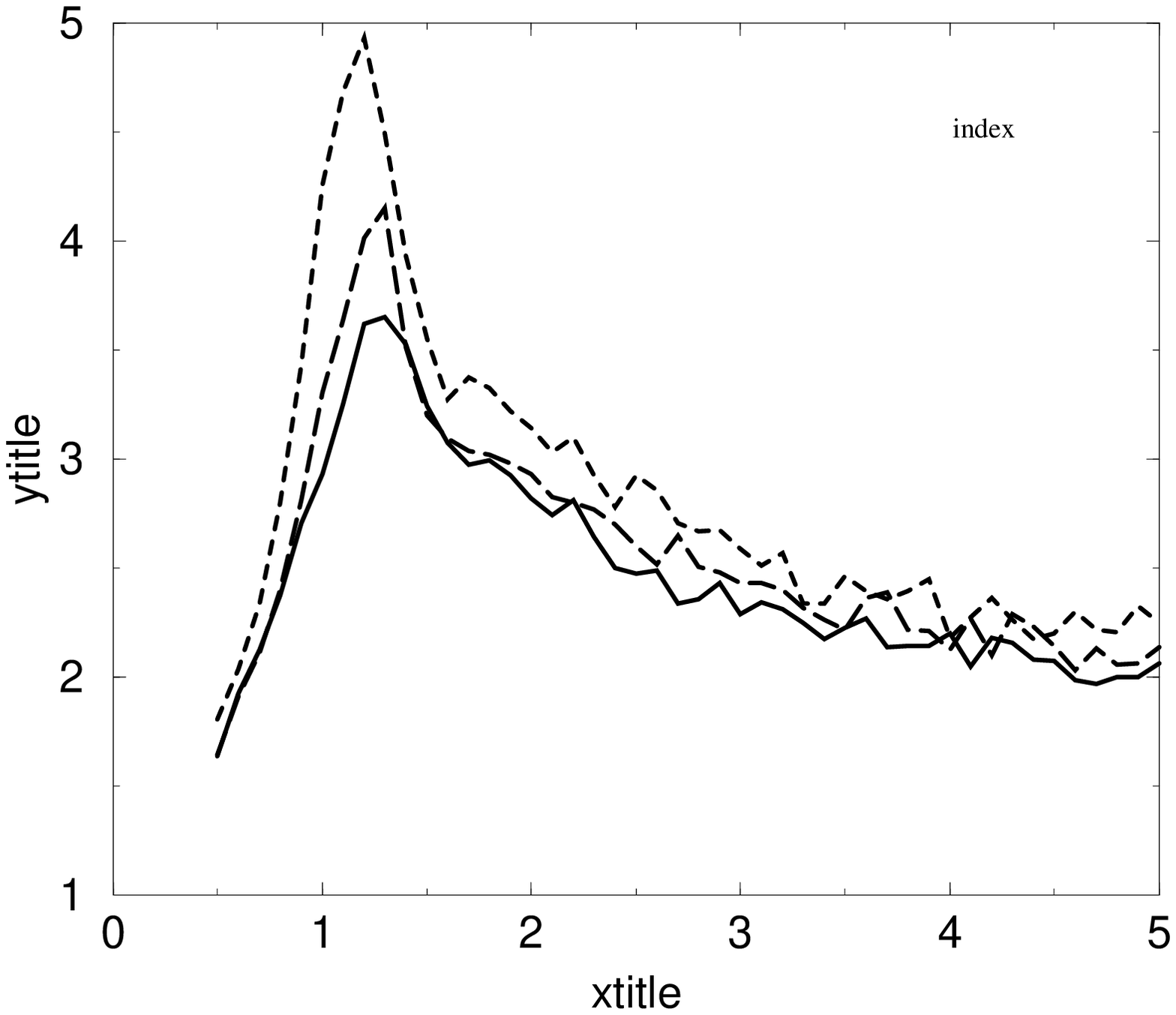,width=0.7\columnwidth}
\caption{Atomic ionization yield for the three frequencies 
$\omega=0.055$ a.u.\ (solid line), $\omega=0.075$ a.u.\ (long dashed line) 
and $\omega=0.11$ a.u. (dashed line) for
Ne$_{16}$ (a) and  Ar$_{16}$ (b). The pulse length was $T=55$ fs.}
\label{Ne.Ar.freq.charge}
\end{center}
\end{figure}

One characteristic feature of the enhanced ionization mechanism is its
relative insensitivity on the frequency of the applied laser field. As long as
the quasistatic picture is valid, the value of $R^*$ should not change
significantly with the laser frequency. On the other hand, any resonance-type
mechanism like the plasmon picture should exhibit a strong dependence of
$R^*$ on the laser frequency. 
We have calculated the ionization yield of Ne$_{16}$ and Ar$_{16}$ for three
different frequencies, with the results shown in \fig{Ne.Ar.freq.charge}.
The position of $R^*$ does hardly change for different laser
frequencies.  Hence, we can exclude any kind of resonance behavior in
favor of the enhanced ionization mechanism.

Although the position of $R^*$ does not change with the laser frequency,
the ionization yield does. This is due to the fact that electrons which are
already outer ionized tend to leave the cluster region faster when
the frequency is smaller: the quiver amplitude of an electron in an electric
field of frequency $\omega$ is proportional to $1/\omega^2$. Hence, on 
average in fields
of higher frequencies the already ionized electrons  will stay
closer to the cluster for a longer time and lead to an increased field
ionization rate.

\section{Exploration of the parameters  controlling laser-cluster 
interaction}
Having established the basic mechanism for coupling energy from the 
laser pulse into small rare gas clusters, we will explore now the 
influence of different parameters on this mechanism, such as cluster 
size, energy content of the laser pulse and laser polarization.
\subsection{Different cluster sizes}
\label{size}
\begin{figure}
\begin{center}
\psfrag{xtitle}[][][0.9]{$R/R_0$}
\psfrag{ytitle}[][][0.9]{atomic charge}
\psfrag{kurve1}[l][][0.9]{Ne$_{16}$}
\psfrag{kurve2}[l][][0.9]{Ne$_{20}$}
\psfrag{kurve3}[l][][0.9]{Ne$_{25}$}
\psfrag{kurve4}[l][][0.9]{Ne$_{30}$}
\epsfig{file=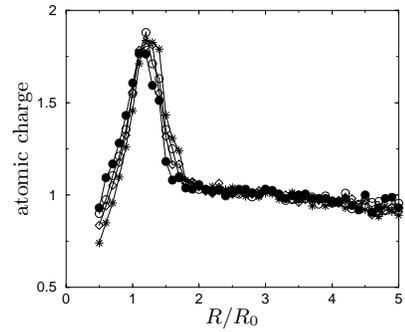,width=0.6\columnwidth}
\caption{Average atomic charge state as a function of the mean interionic
distance for Ne$_{16}$ ($\bullet$), Ne$_{20}$ ($\circ$), Ne$_{25}$ ($\diamond$) 
and
Ne$_{30}$ ($\star$).}
\label{Ne.average_charge.mehrere}
\end{center}
\end{figure}

\begin{figure}
\begin{center}
\psfrag{xtitle}[][][0.9]{$R/R_0$}
\psfrag{ytitle}[][][0.9]{absorbed energy per atom [eV]}
\psfrag{kurve1}[l][][0.9]{Ne$_{16}$}
\psfrag{kurve2}[l][][0.9]{Ne$_{20}$}
\psfrag{kurve3}[l][][0.9]{Ne$_{25}$}
\psfrag{kurve4}[l][][0.9]{Ne$_{30}$}
\epsfig{file=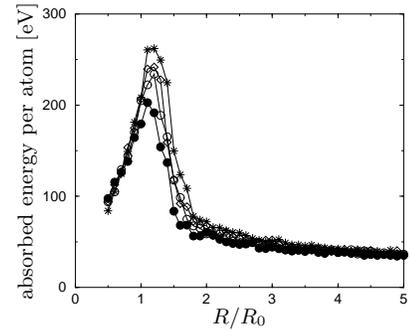,width=0.6\columnwidth}
\caption{Absorbed energy/atom as a function of the mean interionic
distance for Ne$_{16}$ ($\bullet$), Ne$_{20}$ ($\circ$), Ne$_{25}$ ($\diamond$) 
and
Ne$_{30}$ ($\star$).}
\label{Ne.average_energ.mehrere}
\end{center}
\end{figure}

As in the previous section, we present the cluster response with fixed
nuclei first and relate the results of these calculations to the
absorption behavior when the nuclei are allowed to move.  The average
atomic charge and the absorbed energy were calculated as a
function of the mean interionic distance again, analogous to section
\ref{static}.  The equilibrium value $R_0$ does hardly change when
going to bigger clusters.  The variation of $R_{0}$ for Ne$_{16}$,
Ne$_{20}$, Ne$_{25}$ and Ne$_{30}$ is only about 0.01 a.u..  We have
used the same pulse as in section \ref{static}.  As can be seen from
\fig{Ne.average_charge.mehrere} and \fig{Ne.average_energ.mehrere},
the bigger clusters show almost no difference compared to Ne$_{16}$
when the observables are normalized on the number of cluster atoms. 
In particular the existence of a critical distance $R^*>R_0$ is
confirmed in all cases.

There is no hint on a transition to a collective behavior at these cluster
sizes. If we think of cluster physics as the transition regime between atomic
and solid state physics, we are still on the atomic side with a cluster of 30
atoms. 

From the fact that the charge per atom is almost independent of the number of
cluster atoms we may conclude that only the next-neighbor-atoms participate in the
mechanism of enhanced ionization; otherwise the effectivity of this mechanism
should change with the cluster size. The absorbed energy per atom, however, is
varying with the number of atoms. This effect can be easily explained by
calculating the change in the potential energy $U(N)$ of a cluster consisting of
$N$ ions of charge $Z$  and radius $R$ if one adds a new ion with
the same charge $Z$ to the cluster. If one assumes that this new ion is placed
at the border of the cluster, then 
\begin{align}
U(N+1)=U(N)+{NZ^2}/{R}.
\end{align}
If $4\pi r_s^3$ denotes the volume per atom, then $R=N^{1/3}r_s$ and 
\begin{align}
U(N+1)=U(N)+{N^{2/3}Z^2}/{r_s}.
\end{align}
With $N$ as a continuous variable one is left  with the differential
equation
\begin{align}
\frac{dU(N)}{dN}=\frac{N^{2/3}Z^2}{r_s},
\end{align}
so that finally 
\begin{align} 
U(N)=\frac{3}{5}\frac{N^{5/3}Z^2}{r_s}.
\end{align}
Hence, the potential energy per atom $U/N$ has increases with $N^{2/3}$ if the
charge per atom is independent of $N$.

\begin{figure}
\begin{center}
\psfrag{xtitle}[][][0.9]{$R/R_0$}
\psfrag{ytitle}[][][0.9]{atomic charge}
\epsfig{file=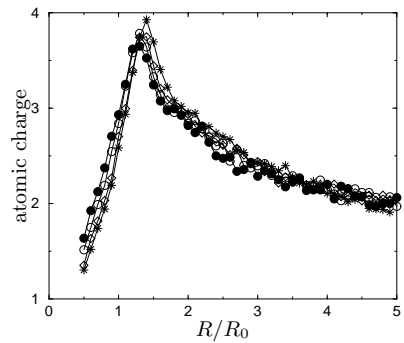,width=0.6\columnwidth}
\caption{Average atomic charge state as a function of the mean interionic
distance for Ar$_{16}$ ($\bullet$), Ar$_{20}$ ($\circ$), Ar$_{25}$ ($\diamond$) 
and
Ar$_{30}$ ($\star$).}
\label{Ar.average_charge.mehrere}
\end{center}
\end{figure}

\begin{figure}
\begin{center}
\psfrag{xtitle}[][][0.9]{$R/R_0$}
\psfrag{ytitle}[][][0.9]{absorbed energy per atom [eV]}
\epsfig{file=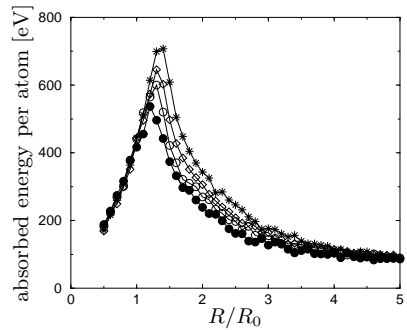,width=0.6\columnwidth}
\caption{Absorbed energy/atom as a function of the mean interionic
distance for Ar$_{16}$ ($\bullet$), Ar$_{20}$ ($\circ$), Ar$_{25}$ ($\diamond$) 
and
Ar$_{30}$ ($\star$).}
\label{Ar.average_energ.mehrere}
\end{center}
\end{figure}

Proceeding from Ne to Ar clusters, one finds again that the
effectivity of the ionization mechanism hardly changes when changing
the cluster size, while the absorbed energy per atom increases with
$N$, for the same reason as discussed above (see
\fig{Ar.average_charge.mehrere} and \fig{Ar.average_energ.mehrere}). 
However, while the Ne clusters show only a little shift of $R^*$ as a
function of cluster size, the ratio of $R^*$ to $R_0$ increases
slightly more with increasing $N$ for Ar.  This is probably due to a 
larger
downshift of the atomic energy levels by the increased total amount of
surrounding charge when $N$ is increased.  As can be seen from
\fig{tunnelh2sketch}, a downshift of the atomic energy levels leads to
an increase in $R^*$.  Since the electron release in argon clusters is 
larger than in Ne clusters higher charged  ions are generated than in
 neon clusters rendering this effect  more pronounced for 
Ar clusters.

\begin{figure}
\psfrag{xtitle}[][][0.9]{T[a.u]}
\psfrag{ytitle}[][][0.9]{cluster charge}
\begin{center}
\epsfig{file=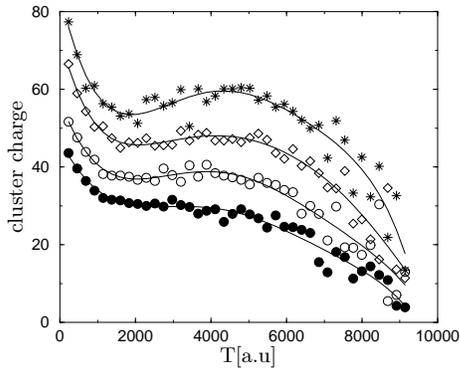,width=0.7\columnwidth}
\caption{Cluster charge as a function of pulse length for 
Ne$_{16}$ ($\bullet$), Ne$_{20}$ ($\circ$), Ne$_{25}$ ($\diamond$) and
Ne$_{30}$ ($\star$). Lines are to guide the eye.} 
\label{Ne.meiwes.charge.mehrere}
\end{center}
\end{figure}

\begin{figure}
\psfrag{xtitle}[][][0.9]{T[a.u]}
\psfrag{ytitle}[][][0.9]{cluster charge}
\begin{center}
\epsfig{file=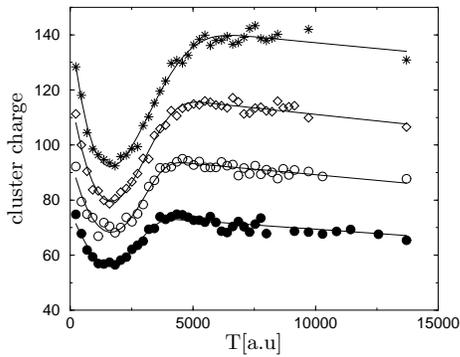,width=0.7\columnwidth}
\caption{Cluster charge as a function of pulse length for
Ar$_{16}$ ($\bullet$), Ar$_{20}$ ($\circ$), Ar$_{25}$ ($\diamond$) and
Ar$_{30}$ ($\star$). Lines are to guide the eye.} 
\label{Ar.meiwes.charge.mehrere}
\end{center}
\end{figure}

Since we have found a critical cluster radius $R^*$ with
$R^*>R_0$ in all cases considered, it is not too surprising that we
find a behavior analogous to the small clusters of 
\fig{meiwes.en} and \fig{meiwes.charge} if 
 the bigger clusters are allowed to expand freely. The results of these
calculations, with the pulse normalization being identical to section
\ref{meiwes}, are shown in \fig{Ne.meiwes.charge.mehrere} and
\fig{Ar.meiwes.charge.mehrere}. For clarity, we have plotted the
total cluster charge instead of the average atomic charge, which is almost the
same independent of $N$.

The overall structure of the curves is seen to be quite similar
throughout the different cluster sizes, as one would expect from
\fig{Ne.average_charge.mehrere} and \fig{Ar.average_charge.mehrere}. 
In the case of Ne clusters, the plateau which has been observed for
Ne$_{16}$ already in \fig{meiwes.charge} goes over into a small maximum
with increasing $N$, which indicates that the enhanced ionization
mechanism is slightly more efficient for larger clusters when the ions
are allowed to move.  One tendency which can be observed for the Ar
clusters is that $T^*$ increases with increasing $N$.  We have seen in
\fig{Ar.average_charge.mehrere} that $R^*$ increases also with $N$, so
that the larger Ar clusters have to travel a longer distance until
they reach the critical radius.  For the Ne clusters, on the contrary,
the curves show almost no shift in the $T$-direction when $N$ is
changed.  We will investigate the dependence of the expansion process
on the various cluster parameters like size and atom charge in closer
detail in section \ref{exp_modell}.

\subsection{The influence of the pulse normalization}
\label{norm}
\begin{figure}
\begin{center}
\psfrag{xtitle}[][][1]{$R/R_0$}
\psfrag{ytitle}[][][1]{atomic charge}
\psfrag{index}[][][1]{a)}
\epsfig{file=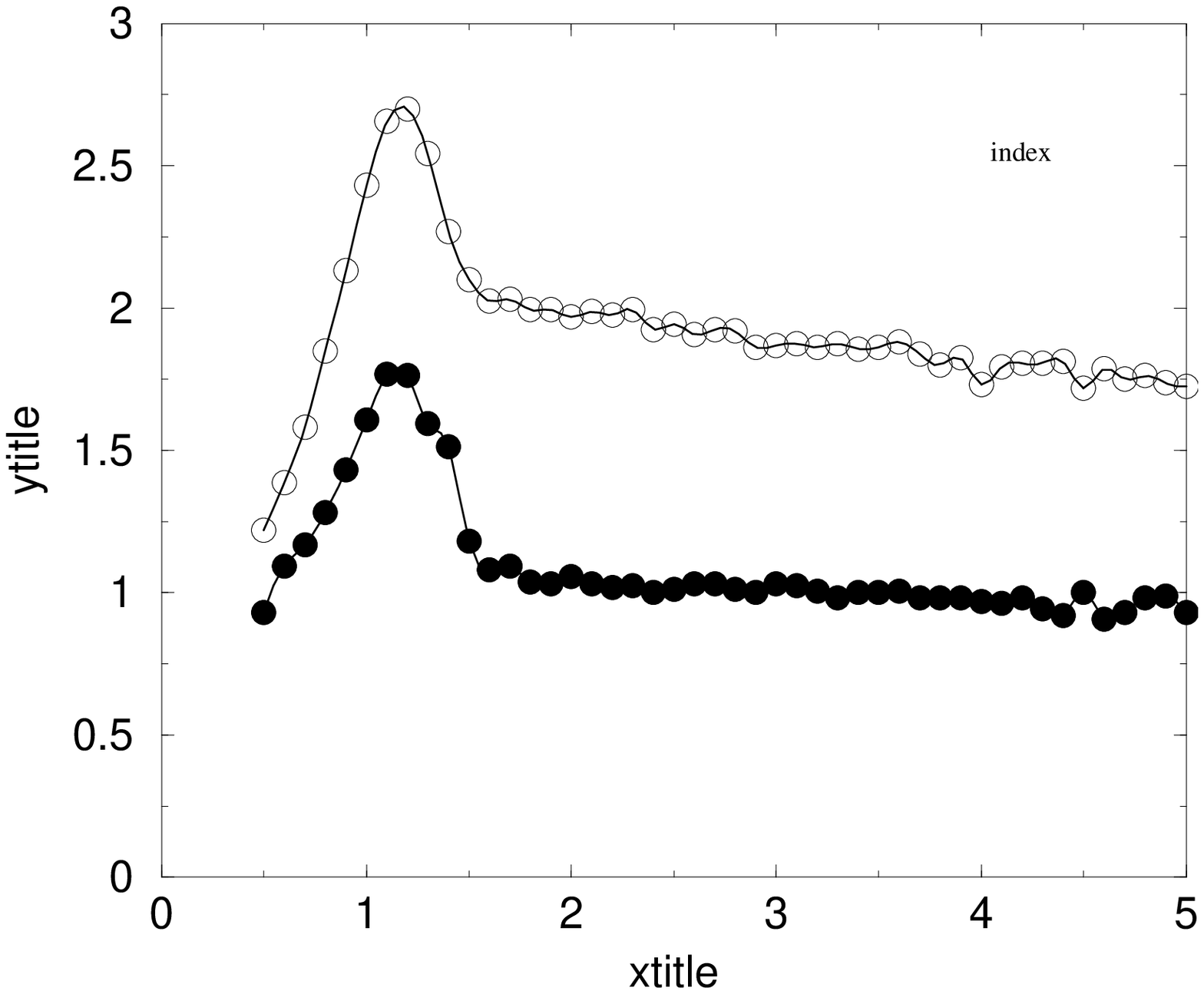,width=0.7\columnwidth}
\psfrag{index}[][][1]{b)}
\epsfig{file=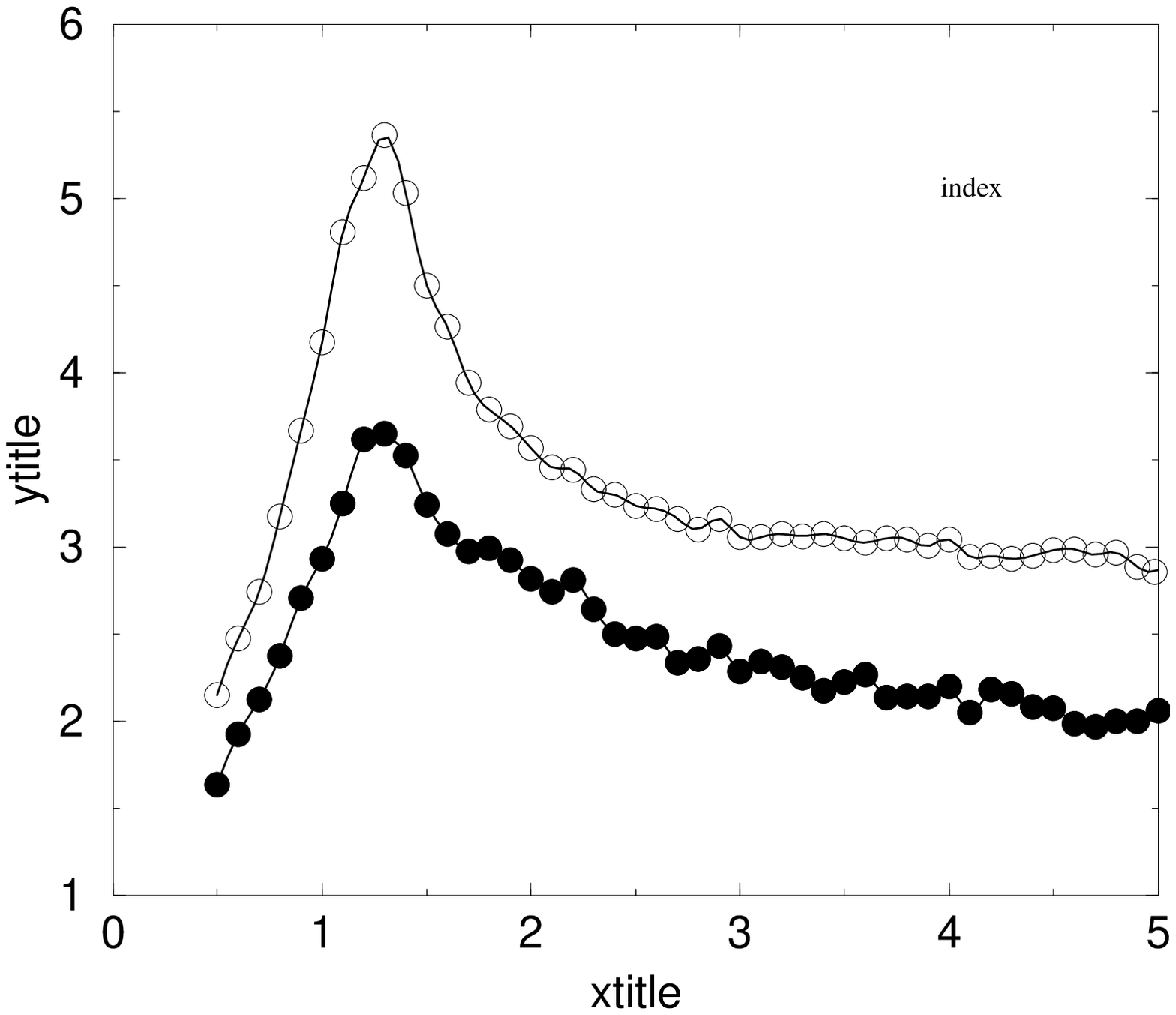,width=0.7\columnwidth}
\psfrag{index}[][][1]{c)}
\epsfig{file=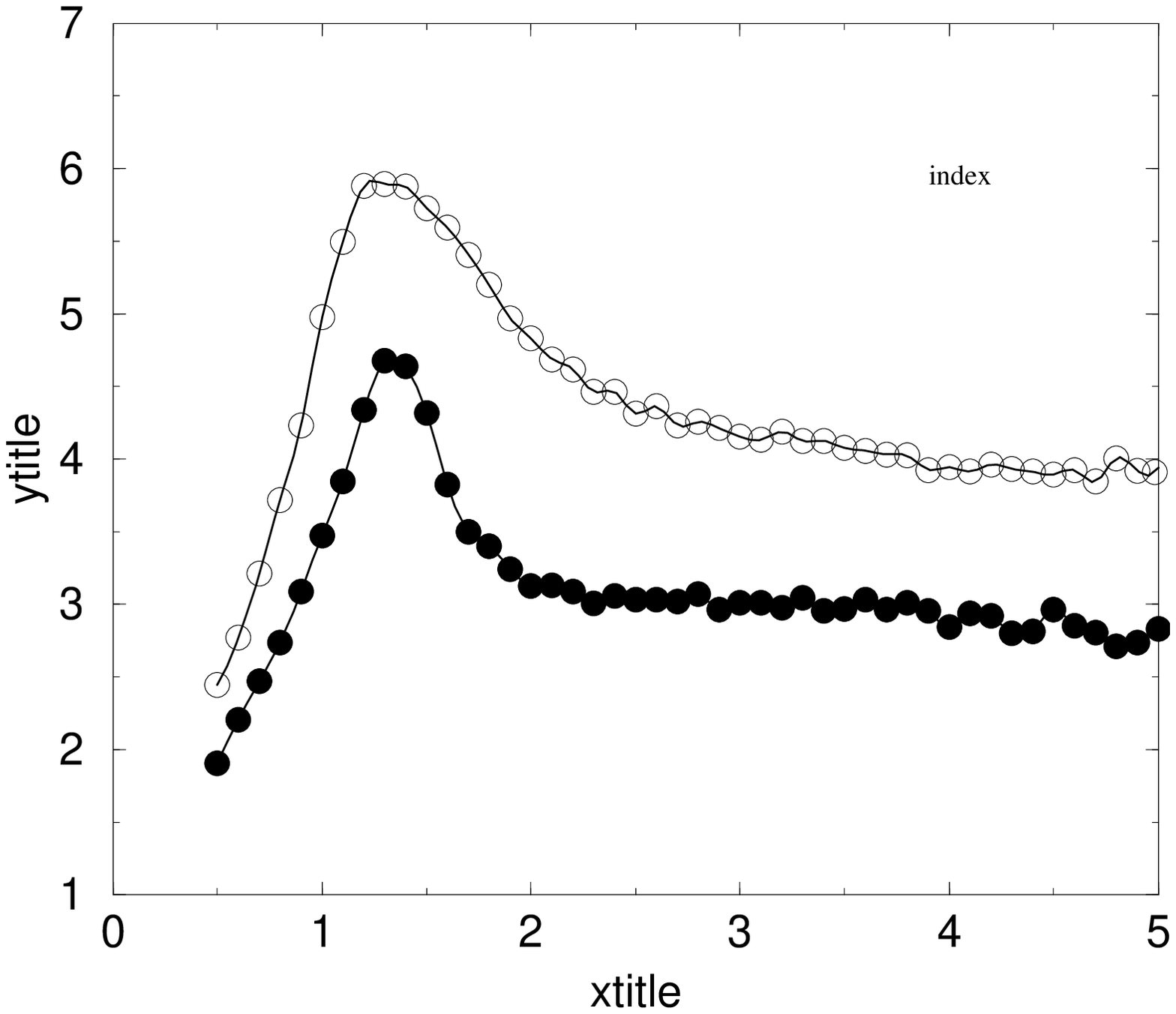,width=0.7\columnwidth}
\psfrag{index}[][][1]{d)}
\epsfig{file=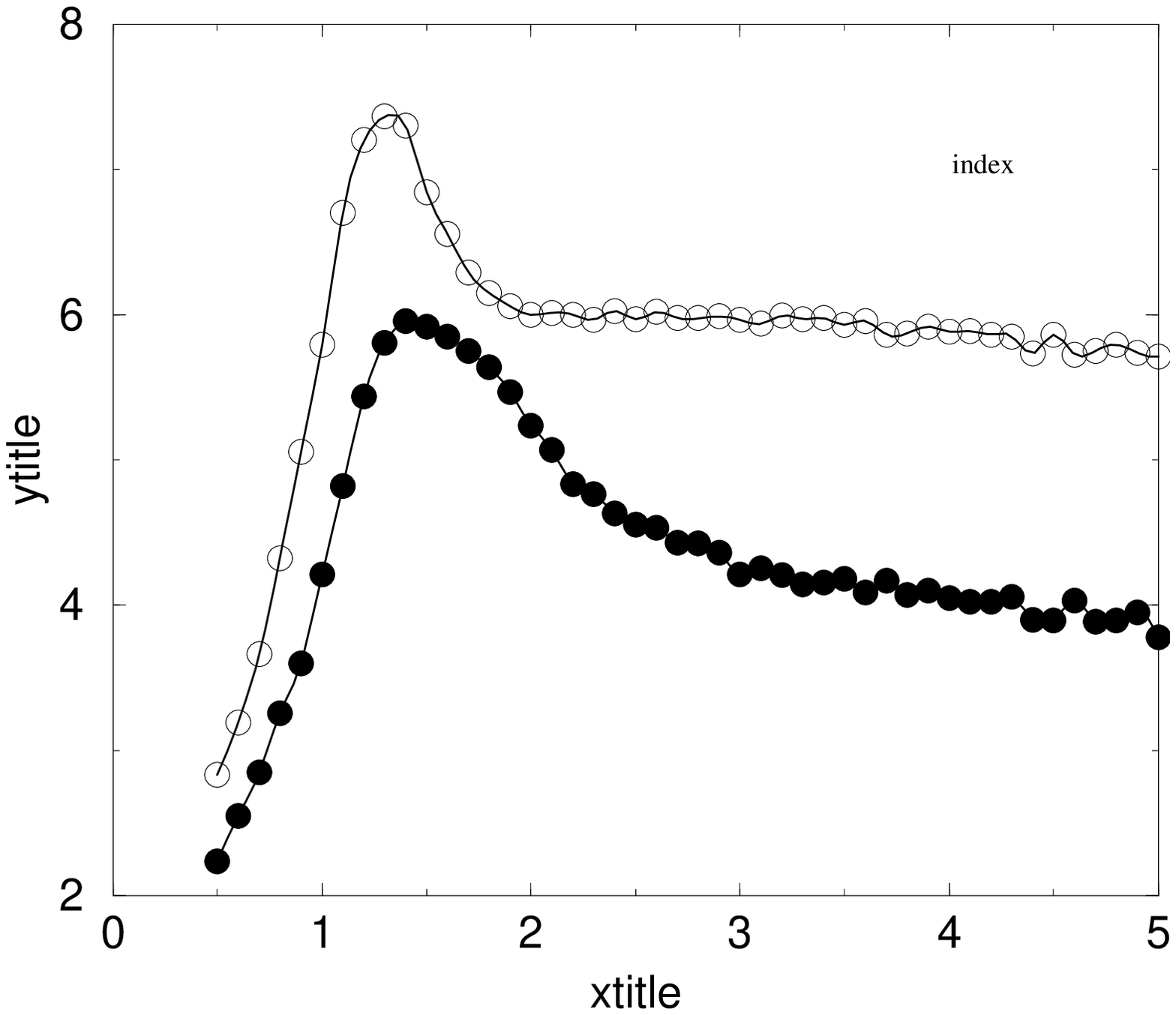,width=0.7\columnwidth}
\caption{Static ionization yield at the  two intensities $
I_1=8.99\cdot 10^{14} \rm{W/cm}^2 (\bullet) and  I_2=2.19\cdot 10^{15}
\rm{W/cm}^2 (\circ)$ for   Ne$_{16}$ (a),  Ar$_{16}$ (b), Kr$_{16}$
(c) and Xe$_{16}$ (d). Lines are to guide the eye.}
\label{intensity.static}
\end{center}
\end{figure}

Changing the laser intensity $I$ in the case of H$_2^+$, with just a single
electron available, leads to a decrease of R$^*$ when $I$ is
increased and vice versa \cite{post1}. For clusters the situation is much more
complicated because with increasing $I$ lower lying energy levels will be
ionized, so that it is a priori not clear in which way a change of the laser
intensity (in a calculation with fixed nuclei) will influence the value of
$R^*$. 
\fig{intensity.static} shows the static ionization yields for Ne$_{16}$,
Ar$_{16}$, Kr$_{16}$ and Xe$_{16}$ under the influence of the pulse used so
far (i.e. a peak intensity of $I_1=8.99\cdot 10^{14} \rm{W/cm}^2$), compared to
the result of a calculation with $I_2=2.19\cdot 10^{15} \rm{W/cm}^2$ (in both
cases the pulse was of the form \eq{Puls} with $\omega=0.055$ a.u. and $T=55$
fs). In all four cases $R^*$ is larger than $R_0$ and can be
reached by cluster expansion. The value of $R^*$ is, if at all, only
slightly decreased in the case of higher intensity: due to the large number of
electrons involved the geometry of the problem is obviously not as sensitive
to the laser field strength as in the H$_2^+$ case.

\begin{figure}
\begin{center}
\psfrag{xtitle}[][][1]{$T$}
\psfrag{ytitle}[][][1]{atomic charge}
\epsfig{file=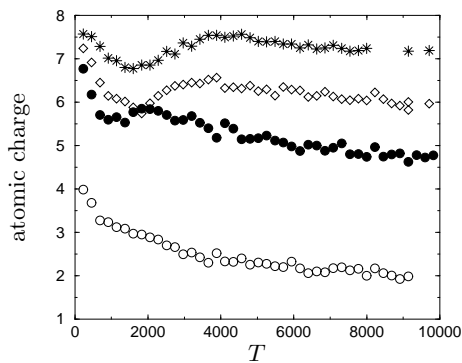,width=0.7\columnwidth}
\caption{Pulse length dependent ionization yields with a pulse energy
corresponding to $I_2=2.19\cdot 10^{15}$ and a pulse length of 20
cycles for Ne$_{16}$ ($\circ$), Ar$_{16}$ ($\bullet$),
Kr$_{16}$ ($\diamond$) and Xe$_{16}$ ($\star$).}
\label{meiwes.intensity}
\end{center}
\end{figure}

Of course, the ionization yield is higher when the intensity is
increased. This leads to significantly shorter expansion times when the nuclei
are allowed to move. Consequently, the optimal pulse lengths $T^{*}$ are now shifted
towards smaller values, as can be seen in \fig{meiwes.intensity} in 
accordance with our picture of the ionization process.

\subsection{Enhanced ionization and circular polarization}
\label{circ}
So far all the results presented are expected to hold also for diatomic
molecules. One main difference between such a molecule and a cluster is the
molecular axis: the whole picture of ENIO as sketched in \fig{tunnelh2sketch}
relies on the fact that the polarization direction of the applied laser field
coincides with the internuclear axis. And indeed, experiments as well as
calculations with a
polarization axis perpendicular to the molecular axis have shown no signature
of enhanced ionization \cite{ban2,ban3}. For the same reason ENIO is 
much less efficient under circular polarization.

\begin{figure}
\begin{center}
\psfrag{xtitle}[][][1]{$R/R_0$}
\psfrag{ytitle}[][][1]{atomic charge}
\epsfig{file=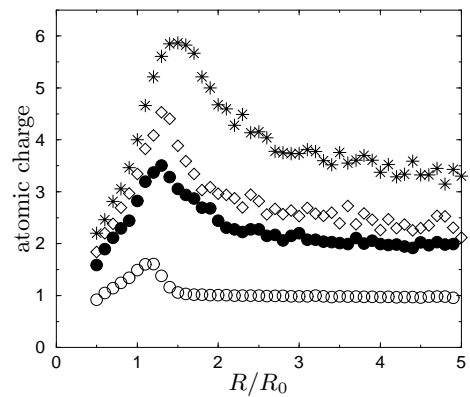,width=0.7\columnwidth}
\caption{Atomic charges with fixed nuclei and circular polarization for
Ne$_{16}$ ($\circ$), Ar$_{16}$ ($\bullet$),
Kr$_{16}$ ($\diamond$) and Xe$_{16}$ ($\star$).
The pulse
parameters are $F_0=0.16/\sqrt{2}$ a.u., $\omega=0.055$ a.u.\ and $T=55$ fs.}
\label{static.charge.circ}
\end{center}
\end{figure}

\begin{figure}
\begin{center}
\psfrag{xtitle}[][][1]{$T$ [a.u.]}
\psfrag{ytitle}[][][1]{atomic charge}
\epsfig{file=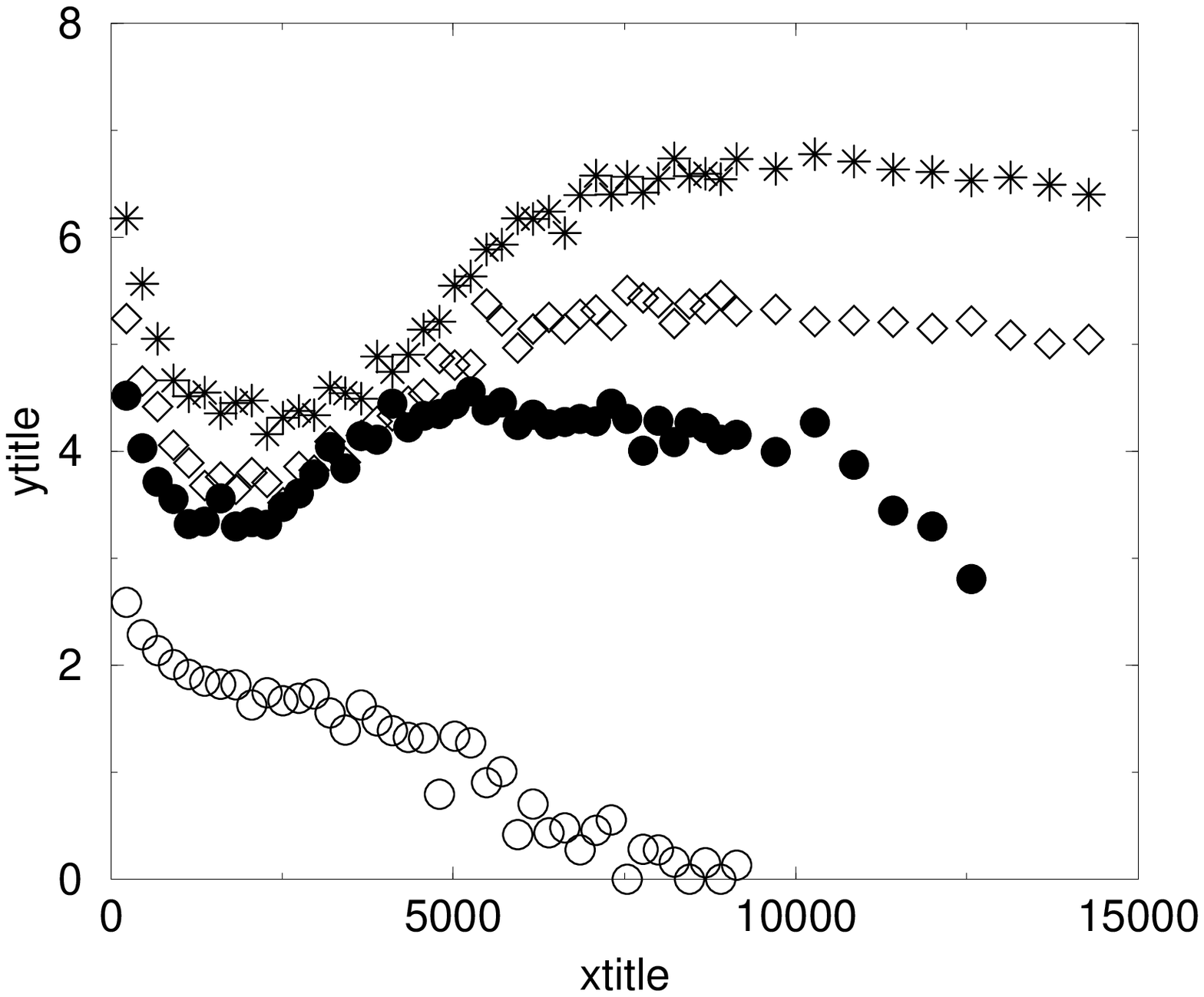,width=0.7\columnwidth}
\caption{Same as \fig{static.charge.circ} but 
with moving nuclei.}
\label{meiwes.charge.circ}
\end{center}
\end{figure}

On the other hand, a cluster is (in first approximation) spherically
symmetric. Thus one would expect enhanced ionization to work also with
circularly polarized light. To test this hypothesis, we have performed the same
calculations as in the previous sections, but now with circularly polarized
laser light. We have chosen the field strength of the laser
such that the energy content of a pulse with a certain pulse length $T$ remains
constant when passing from linear to circular polarization: if 
\begin{align}
F_0\sin^2(\pi/Tt)\sin(\omega t)\vec{e}_x
\end{align} 
is a laser pulse with linear polarization in $x$-direction, then
\begin{align}
F_0/\sqrt{2} \sin^2(\pi/Tt)\left(\sin(\omega t)\vec{e}_x+\cos(\omega
t)\vec{e}_y\right)
\end{align}
is the corresponding circularly polarized pulse.  With this definition
the maximum field strength is decreased by a factor of $\sqrt{2}$.  As
expected, ENIO also exists for circularly polarized laser pulses. 
\fig{static.charge.circ} shows the calculations with fixed nuclei,
\fig{meiwes.charge.circ} the corresponding results with moving nuclei. 
In the case of static nuclei, we find that the ionization yield in the
critical regime is almost as high for circular as for linear
polarization (\fig{static.charge}), in sharp contrast to the
above-mentioned results for diatomic molecules.  Consequently, when
the nuclei are allowed to move we also get qualitatively the same
results (\fig{meiwes.charge.circ}) as with linear polarization
(\fig{meiwes.charge}).  It is only for rather long pulses
that in the case of Ne$_{16}$ and Ar$_{16}$ the ionization yield is
significantly lower than in the linear case, which is due to the
reduced maximum field strength. 
Summarizing our exploration of different parameters we find that ENIO 
for clusters is a rather robust phenomenon. This has motivated us to 
 ask if the optimum pulse length $T^{*}$ can be quantitatively linked
to the critical radius $R^{*}$. 

\section{Analytical formula for the Coulomb explosion}
\label{exp_modell}
To isolate the relation of $T^{*}$ to $R^{*}$ we divide the 
time-dependent dynamics into 
 three different phases: phase I denotes
the time from the onset of the laser pulse until 50\% of the 
atoms in the cluster have lost one  electron due to inner ionization.
We will refer to this time as $T_0$ subsequently. Since
some of the inner ionized electrons will 
leave the cluster, we can say that $T_0$ marks the beginning of the expansion
process.

In this first phase inner ionization is dominated by atomic processes,
the environment plays only a minor role.   
For a single atom/ion the time-dependent probability that the active
electron is {\em not yet} ionized reads, in terms of the field- and
binding energy dependent ionization rate $w(f(t),E_b)$ 
\begin{align}
P_{\rm{neutral}}(t)=\exp\left(-\int_0^t w(f(t'),E_b) \ dt'\right),
\end{align} 
where $E_b$ denotes the binding energy.  The probability
that in a cluster consisting of $N$ such atoms no electron has been
ionized is given by 
\begin{align}
P_{\rm{neutral}}^{\rm{cluster}}(t)=[P_{\rm{neutral}}(t)]^N.
\end{align} 
The exponential dependence on $N$ renders
$P_{\rm{neutral}}^{\rm{cluster}}(t)$ practically a step function. 
Hence, the exact value (between zero and one) for the definition of
$T_0$ is not relevant.  We determine $T_{0}$ from
$P_{\rm{neutral}}^{\rm{cluster}}(T_{0}) =1/2$ which is tantamount to
demanding that on average 50 \% of the atoms in the cluster are singly
ionized at $T_{0}$.

\begin{figure}
\centering
\psfrag{xtitle}[t][][1.1]{$t$}
\psfrag{ytitle}[][][1]{$R(t)$}
\psfrag{t1}[][][0.8]{I}
\psfrag{t2}[][][0.8]{II}
\psfrag{t3}[][][0.8]{III}
\psfrag{t4}[][][0.8]{$T_0$}
\psfrag{t5}[][][0.8]{$T$}
\psfrag{t6}[l][][0.8]{$R^*$}
\epsfig{file=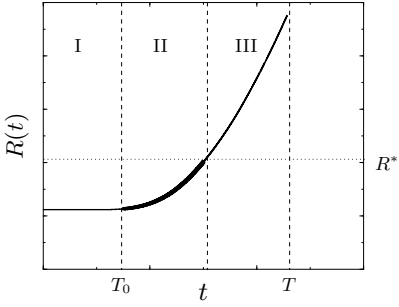,width=0.6\columnwidth}
\caption{Sketch of phases I, II and III during the pulse (see text)}
\label{phases}
\end{figure}

The second phase contains the  cluster expansion  up to the critical time
$T^*$, when the critical cluster radius $R^*$ is reached. Hence, the critical
time is the sum of $T_0$ and the expansion time $T_{\rm{exp}}$:
\begin{align}
T^*=T_0+T_{\rm{exp}}.
\end{align}
The third phase is finally the time from reaching $R^*$ to the end
of the pulse. A schematic picture of the different phases is shown in
\fig{phases}.

From $T_0$ until the end of the pulse, the total cluster charge increases from
$Z=N/2$ to $Z=Z_{\rm final}$. As a first approximation, we assume that the expansion
from $R=R_0$ to $R=R^*$ is driven by an effective charge per atom
$\bar{Z}=\alpha Z_{\rm final}/N$ with a constant factor $\alpha$ 
which stands for
the efficiency of the enhanced ionization mechanism. Furthermore, the
expansion is assumed to be accomplished by the Coulomb repulsion of the nuclei
only, i.e. we neglect the influence of the laser field as well as of the electronic dynamics on the
expansion process. Under these two assumptions, we can use energy conservation
to write
\begin{align}
\label{toten}
\sum_{i=1}^N\frac{M}{2}
v_i^2+\sum_{(i\neq j)=1}^N\frac{\bar{Z}^2}{R_{ij}(t)}=E,
\end{align}
where  $R_{ij}(t) = |\vec{R}_i(t)-\vec{R}_j(t)|$, $M$ is the atomic mass 
and $v_i$ the respective atomic velocities.
As a further approximation we  assume that the expansion takes place in a
homogenous and isotropic way, so that it can be described by a common
expansion parameter $\lambda(t)$ with
$\vec{R}_i(t)=\lambda(t)\vec{R}_i(0)$.  Defining
\begin{align}
K_0 &:=\sum_{i=1}^N\frac{1}{2}M
R_i^2(0) \nonumber\\
\ V_0& :=\sum_{(i\neq
j)=1}^N\frac{(Z_{\rm final}/N)^2}{R_{ij}(0)}
\end{align}
and taking into account that the kinetic energy is zero before the 
expansion we may write the energy balance of \eq{toten}
as
\begin{align}
\label{lamben}
K_{0}\dot{\lambda}^2(t)+\frac{\alpha^2}{\lambda(t)}V_{0} = 
\frac{\alpha^2}{\lambda(0)}V_{0}\,.
\end{align} 
Finally, \eq{lamben} may be rearranged as a
 differential equation for $\lambda(t)$
\begin{align}
\frac{d\lambda(t)}{dt}=\alpha\left[\left(1-\frac{1}{\lambda(t)}\right)\frac{V_0}{K_0}\right]^{1/2},
\end{align}
which can be solved analytically by separation of  variables for
the expansion time:
\begin{align}
\label{T-Gleichung}
T_{\rm exp}=&\sqrt{\frac{K_0}{V_0}}\frac{1}{\alpha}\left[\sqrt{x(x-1)}+\log\left(\sqrt{x-1}+\sqrt{x}\right)\right]_{x=1}^{\lambda}\nonumber\\
=:&\sqrt{\frac{K_0}{V_0}}\frac{f(\lambda)}{\alpha}\,.
\end{align}
The ratio $K_0/V_0$  determines the time scale for the expansion of the
cluster. By replacing $R_{ij}(0)$, the distance between two
ions in $V_0$, with the cluster radius $R$ (which would be an exact
approximation if all ions were placed on the surface of the cluster), we can
estimate how this time scale depends on the characteristic variables of a cluster:
\begin{align}
\frac{T_0}{V_0}\propto \frac{MR^3}{(N-1)(Z_{\rm final}/N)^2}.
\end{align}  
From this equation we can read off how the expansion process changes when the
number of atoms $N$ is changed while keeping all other parameters fixed: if
$V_{\rm{atom}}=4/3 \pi r_s^3$ is the volume of one atom inside the
cluster, then $R^3=N r_s^3$. Hence, the time scale of the expansion is governed
by the factor $N/(N-1)$, which depends only  weakly on $N$. This leads us
to the conclusion that indeed the increase of $T^*$ with $N$ in
\fig{Ar.meiwes.charge.mehrere} is not due to the expansion process but 
due to the increase of $R^*$ with $N$ only (see \fig{Ar.average_charge.mehrere}).
\subsection{Scaling of the optimal pulse length}
With \eq{T-Gleichung} at hand we are able to set up a relation between the
optimal pulse lengths for various cluster if we make one last assumption: the
factor $\alpha$, which determines the ratio between the average atomic charge
$\bar{Z}$ driving the expansion up to $R^*$ and the final charge per atom
after the pulse, $Z_{\rm{final}}/N$, is identical for all clusters. If this
hypothesis was true, then $1-\alpha$ would be a universal measure for the 
efficiency of the ENIO mechanism.

\begin{figure}
\begin{center}
\psfrag{xtitle}[t][][1]{$(K_{0}/V_{0})^{1/2}f(\lambda)$}
\psfrag{ytitle}[][][1]{$T_{\rm exp}$}
\epsfig{file=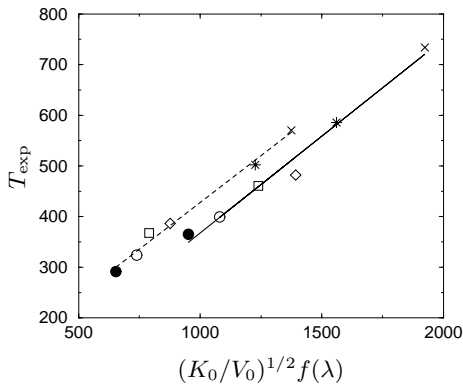,width=0.7\columnwidth}
\caption{Expansion time (numerical data) as a function of
$(K_{0}/V_{0})^{1/2}f(\lambda)$ and linear fits (see text). Two different energy normalizations
were used: $F_0=0.16$ a.u. (solid line) and $F_0=0.25$ a.u. (dashed line), both at a frequency of
$\omega=0.055$ a.u. and a pulse length of $T=55$ fs.   
Ar$_{16} (\bullet)$, Ar$_{20} (\circ)$, Ar$_{25} (\Box)$, Ar$_{30} (\diamond)$,
Kr$_{16} (\star)$ and Xe$_{16} (\times)$}
\label{linear}
\end{center}
\end{figure}

If $\alpha$ is the same for all clusters, we get from \eq{T-Gleichung} a linear relation
between the expansion time $T_{\rm exp}$ and 
$f(\lambda)(T_0/V_0)^{1/2}$, different for each 
cluster.  This prediction is
confirmed by \fig{linear} which shows  the expansion times $T_{\rm 
exp}= T^*/2-T_0$ as a
function of the cluster-dependent values of
$(K_0/V_0)^{1/2}f(\lambda)$ for different clusters.  We have obtained 
$\lambda$ 
from the respective static calculations for each cluster.  A linear
fit to the data yields $\alpha=0.38$ and $\alpha=0.37$ for energy
normalized pulses at $F_0=0.16$ and $F_0=0.25$, respectively.  The
correlation coefficient is in both cases higher than 0.99.  Hence,
$\alpha$ is the same for different clusters, and it is even almost the
same for different energy normalizations of the laser pulse.  This
result {\em a posteriori} justifies the approximations we have made in
establishing our expansion model. 

The fact that $\alpha$ remains almost the same when changing the
pulse normalization is certainly an unexpected result; it is probably
valid only for a limited range of pulse energy contents, if one thinks
of $\alpha$ as a measure for the efficiency of ENIO. At least in the
limit of a very large pulse energy, when the electric field of the
laser is larger than the electric field from the charges in the
cluster, we expect the ENIO mechanism to play no important role any
more, since the cluster geometry will be washed out.  However,  the
good agreement of the linear fit in \fig{linear} with our numerical
data for each of the two normalizations separately points to a deeper
scaling relation between the various clusters, the reason for which
will be explored in future work.

\section{Summary and conclusion}
We have developed a quasiclassical model for small rare gas cluster in strong
laser fields. This model allows us to investigate the influence of 
several parameters, like the atomic element and the
cluster size, but also the characteristica of the applied laser field. We have
shown that, as a function of pulse length, the energy absorption as well as
the ionization yield of all but the Ne clusters show a clear maximum when the
energy content of the pulse is kept fixed. This behavior has been attributed
to the existence of a critical cluster radius $R^*$, whose origin could be
explained by generalizing the CREI or ENIO concept from diatomic molecules to
small rare gas clusters. It was shown that this mechanism is stable against a
change of system parameters, even when switching from linear to circular
polarization. This is a pronounced difference between clusters and molecules.

Finally, we were able to condense the absorption and expansion process into a
 simple model and obtained an analytical expression connecting the expansion
time and the cluster properties. The validity of this expression has been
confirmed by our numerical data.

Future investigations will include the transition to the plasma regime, which
we expect to begin at around $N\approx10^3$. In this regime, another critical
radius should emerge, when the laser frequency matches the plasma frequency. It
will be interesting to see how this additional critical radius evolves with
$N$ and if there is a regime where the $R^*$ coming from the ENIO process and
the plasma related $R^*_{c}$ coexist.

We would like the DFG for financial support within the Gerhard 
Hess-program.

\bibliography{clusterpaper-eprint}

\begin{thebibliography}{27}
\bibitem[{foo({\natexlab{a}})}]{footnote1}
\bibinfo{note}{From this moment on, we neglect the contribution of the
  Lennard-Jones potential.}
\bibitem[{foo({\natexlab{b}})}]{footnote2}
\bibinfo{note}{Energy conservation is not automatically guaranteed by just
  placing the electron at $r_2$ with no kinetic energy, as one might suspect
  from \fig{tunnelbild}. The reason is that the ionic softcore parameter is
  $Z$-dependent and, in addition, differs from the softcore parameter of an
  electron.}
\end{thebibliography}

\end{document}